# Extrinsic Limitations of Stealthy Hyperuniform devices

Yuhao Xu[1], Miao Chen[1], Louis Forestier[1], Franck Carcenac[2], Laurent Mazenq[2], and Philippe Lalanne[1*]

[1]Laboratoire Photonique Numérique et Nanosciences (LP2N), Université de Bordeaux, Institut d'Optique Graduate School, CNRS, Talence, France

[2]LAAS-CNRS, Université de Toulouse, CNRS, Toulouse, France

[*]Corresponding author: Philippe.Lalanne@institutoptique.fr

**Abstract**: Hyperuniformity promises an unusual form of wave control: the suppression of elastic scattering over extended angular ranges without periodic order. Here, we present a comprehensive experimental and theoretical study of 2D stealthy hyperuniform metasurfaces operating at optical frequencies. In agreement with theoretical expectations, we observe a pronounced reduction of elastic scattering around the specular direction in metasurfaces fabricated by electron-beam lithography. However, the measured suppression is substantially weaker than that predicted by structure-factor calculations based on ideal stealthy hyperuniform point-pattern generators. We identify and quantitatively analyze the physical origins of this discrepancy and establish realistic performance bounds. By isolating the dominant limiting mechanisms, our results provide practical design guidelines for the implementation of stealthy hyperuniformity in functional devices.

## 1. Introduction

The study of light propagation in complex media has been an active field for several decades, stimulating fundamental questions in mesoscopic physics [1] and enabling a wide range of photonic applications [2]. More recently, the ability to engineer disorder through controlled structural correlations has opened new avenues for designing materials with tailored optical responses [3-9].

Hyperuniformity [10-13] is a general concept introduced to characterize a class of correlated disordered systems that exhibit a pronounced suppression of density fluctuations at large length scales, while remaining disordered at short scales. This concept has found applications across diverse fields, ranging from photonic materials [14,15], as considered here, to biological systems [16] and cosmology [17]. For discrete systems composed of identical particles, hyperuniformity is characterized by the structure factor, which depends only on the relative particle positions. For systems that can be represented as weighted point patterns, a weighted structure factor, in which each point carries an individual weight, provides the appropriate measure of hyperuniformity [18,19]. Formally, hyperuniform point patterns are characterized by structure factors $S_r$ that vanish as the wavelength tends to infinity, $S_r(\mathbf{q}) \to 0$ as the reciprocal wavevector $\mathbf{q}$ tends towards zero.

In this work, we investigate *stealthy hyperuniform* (SHU) structures, a particular class of hyperuniform systems in which density fluctuations are completely suppressed over a finite region of reciprocal space, such that $S_r(\mathbf{q}) = 0$ for all $|\mathbf{q}|$ smaller than a prescribed cutoff $q_{\max}$.

We focus on two-dimensional (2D) hyperuniform systems rather than their three-dimensional counterparts because planar geometries are naturally compatible with top-down nanofabrication techniques and enable precise control over the positions of the constituent scatterers.

Consistently, throughout this work, $\mathbf{q}$ denotes the in-plane wavevector component $\mathbf{q}_{\parallel} = [q_x, q_y]$ parallel to the surface.

We further consider out-of-plane diffraction rather than in-plane transport phenomena [20-24], a regime that has received comparatively limited experimental attention despite previous studies in other wave systems such as acoustic and radar waves [25-27]. Consistent with this geometry, we refer to the structures under investigation as hyperuniform metasurfaces [28] and to the individual scatterers as meta-atoms. Although our analysis focuses on out-of-plane scattering, we expect the main conclusions regarding the fundamental limitations to remain qualitatively valid for in-plane transport.

Within scattering theory, $\mathbf{q}_{\parallel}$ corresponds to the in-plane momentum transfer between the scattered and incident plane waves and stealthy hyperuniformity is expected to strongly inhibit elastic scattering over a well-defined angular range around the specular light direction $\mathbf{q}_{\parallel} = \mathbf{0}$ (Figure 1). We define the quenching zone as the finite region of momentum space over which this suppression occurs, and the quenching efficiency as the reduction in scattered intensity per solid angle inside the quenching zone relative to that outside it. Together, these quantities provide natural figures of merit for evaluating and optimizing the angular scattering response of hyperuniform metasurfaces.

Several algorithms [20,29-31] have been developed to generate hyperuniform point patterns that exhibit extremely large theoretical quenching efficiencies, often exceeding $10^5$. These algorithms are both effective and flexible, and can be extended to anisotropic configurations with distinct scattering responses along the $q_x$ and $q_y$ directions [32]. The spatial extent of the quenching zone is predicted to depend on parameters such as the wavelength and the meta-atom density through a relation that we term the *quenching equation*, by analogy with the grating equation governing diffraction from periodic surfaces.

Using hyperuniform metasurfaces fabricated by electron-beam lithography, we provide a quantitative experimental validation of this quenching equation and determine the quenching efficiencies attainable under realistic conditions. Our measurements reveal a substantial gap between theory and experiment, with theoretical models significantly overestimating the achievable quenching efficiency.

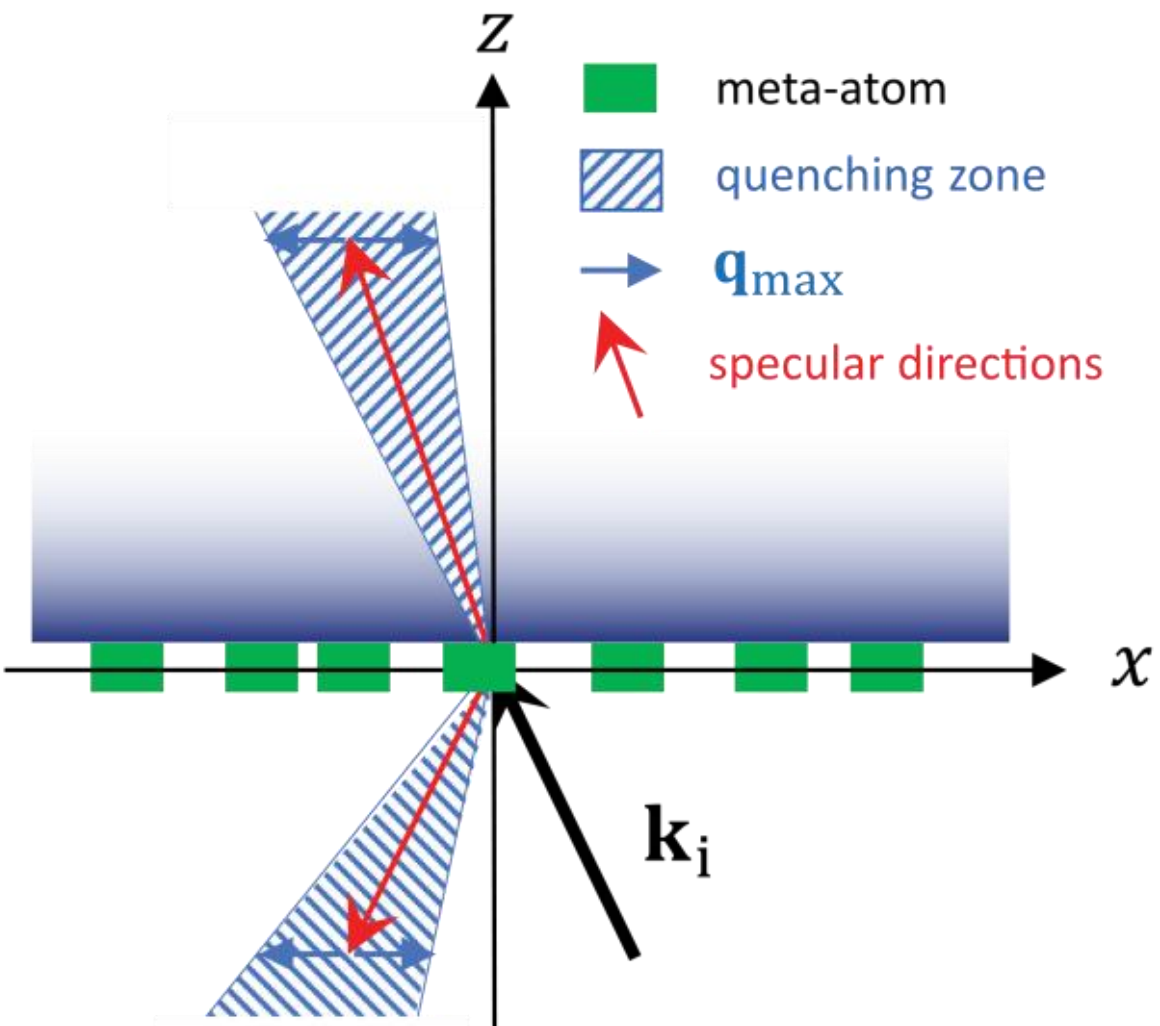

**Figure 1**. Quenching zone of stealthy hyperuniform metasurfaces illuminated by an obliquely incident plane wave with wavevector $\mathbf{k}_i = [\mathbf{k}_{i,\parallel}, k_{i,z}]$. Scattering is suppressed around the specular direction (red arrows), defined by $\mathbf{k}_{s,\parallel} = \mathbf{k}_{i,\parallel}$. Specifically, no light is scattered into directions, whose in-plane wavevectors satisfy $\mathbf{k}_{s,\parallel} = \mathbf{k}_{i,\parallel} + \mathbf{q}_{\parallel}$, with $|\mathbf{q}_{\parallel}| < q_{max}$. The corresponding angular domain defines the quenching zone.

There are several reasons why the strong quenching predicted theoretically by the structure factor may not be observed experimentally. To identify them, let us consider a metasurface composed of a collection of $N$ meta-atoms, located in the plane $z = 0$ at positions $\mathbf{r}_p = [\mathbf{r}_{p,\|}, z_p = 0]$, with $p = 1 \ldots N$. The metasurface is illuminated by a planewave with a wavevector $\mathbf{k}_\mathrm{i} = [\mathbf{k}_{\mathrm{i},\|}, k_{\mathrm{i},z}]$.

We consider the field scattered into a plane wave with wavevector $\mathbf{k}_\mathrm{s} = [\mathbf{k}_{\mathrm{s},\|}, k_{\mathrm{s},z}]$ and define the in-plane momentum transfer $\mathbf{q}_\| = \mathbf{k}_{\mathrm{s},\|} - \mathbf{k}_{\mathrm{i},\|}$. The total scattered electric field can be written as a superposition of the fields, $\mathbf{E}_\mathrm{s}^{(p)}(\mathbf{r})$, scattered by the individual meta-atoms,

$$\mathbf{E}_\mathrm{s}(\mathbf{r}) = \sum_{p=1}^{N} \mathbf{E}_\mathrm{s}^{(p)}(\mathbf{r}) \exp\left[-i\mathbf{q}_\| \cdot \mathbf{r}_{p,\|}\right], \tag{1}$$

where the exponential factor $\exp\left[-i\mathbf{q}_\| \cdot \mathbf{r}_{p,\|}\right]$ is a dephasing term between the incident and scattered planewaves and is introduced for convenience. A common simplifying assumption is that all meta-atoms scatter identically, such that $\mathbf{E}_\mathrm{s}^{(p)} = \mathbf{E}_\mathrm{s}^{(0)}(\mathbf{r})$ for all $p$. Under this assumption, the scattered field factorizes as $\mathbf{E}_\mathrm{s}(\mathbf{r}) = \mathbf{E}_\mathrm{s}^{(0)}(\mathbf{r}) \sum_{p=1}^{N} \exp\left[-i\mathbf{q}_\| \cdot \mathbf{r}_{p,\|}\right]$, and the interference sum, $\sum_{p=1}^{N} \exp\left[-i\mathbf{q}_\| \cdot \mathbf{r}_{p,\|}\right]$, then plays a central role. Upon taking the modulus squared to obtain the scattered intensity, this term gives rise to the structure factor [28],

$$S_\mathrm{r}(\mathbf{q}_\|) = \frac{1}{N} \left|\sum_{p=1}^{N} \exp\left[-i\mathbf{q}_\| \cdot \mathbf{r}_{p,\|}\right]\right|^2. \tag{2}$$

Any departure from the condition $\mathbf{E}_\mathrm{s}^{(p)}(\mathbf{r}) = \mathbf{E}_\mathrm{s}^{(0)}(\mathbf{r})$ for all $p$ prevents the structure factor from being directly imprinted onto the properties of the diffuse scattered light. Several mechanisms can lead to such deviations:

- Polydispersity. If the meta-atoms are not strictly identical, their scattering responses differ from one another, causing the factorization assumption to break down.
- Multiple scattering. Even for perfectly identical meta-atoms, the local excitation field generally varies from site to site because it includes contributions from waves scattered by all other meta-atoms. The factorization assumption is therefore expected to hold only in sufficiently dilute metasurfaces, where multiple scattering remains weak.

Additional limitations arise from finite-size effects imposed by experimental constraints, including the finite spatial coherence of the incident beam and the finite lateral extent of the metasurface. Furthermore, for the periodically tiled supercells considered hereafter, fabrication imperfections such as electron-beam stitching errors may introduce additional scattering and partially obscure the signature of hyperuniformity.

By systematically identifying and quantifying these limiting mechanisms, we establish realistic performance bounds for stealthy hyperuniform metasurfaces and clarify their potential for practical photonic applications.

## 2. Experimental results

**SHU pattern generation.** We adapted the algorithms introduced in [20,32], which generate SHU point patterns by minimizing an effective interaction potential under periodic boundary conditions (Figure 2a). For completeness, we briefly recall the key definitions and quantities used in the pattern-generation procedure.

We consider $N$ points distributed within a square domain of side length $L$ subject to periodic boundary conditions. Owing to periodicity, the allowed wavevectors form a square reciprocal lattice with spacing $\Delta q_x = \Delta q_y = 2\pi/L$, and the structure factor is non-null (and defined) only at these discrete reciprocal-lattice vectors.

The SHU generator optimizes the positions of the $N$ points under the constraint that the structure factor remains nearly zero for all reciprocal-lattice vectors contained within a disk of radius $q_{\max}$ centered at the origin of reciprocal space. The number of constrained wavevectors $M(q_{\max})$ depends on $q_{\max}$ and satisfies $2M(q_{\max}) + 1 = \pi(q_{\max}/2\Delta k)^2$ [32,33].

The degree of order is quantified by the stealthiness parameter $\chi = M(q_{\max})/2N$, which measures the fraction of constrained degrees of freedom. For $\chi = 0$, the point pattern is completely random, whereas increasing $\chi$ progressively enhances positional order. For $\chi > 0.5$, the patterns become increasingly crystal-like, and $\chi = 1$ corresponds to a fully constrained periodic lattice.

Combining the above definitions yields a direct relation between $q_{\max}$ and $\chi$, referred to hereafter as the quenching equation [11]:

$$q_{\max} = \sqrt{(4N\chi + 1)/\pi}\,\Delta k \approx 4\sqrt{\pi\rho\chi}. \tag{3}$$

where $\rho = N/L^2$ is the point density.

Figure 2b shows the structure factor of a periodic SHU pattern with $\chi = 0.35$. Within the disk $|\mathbf{q}| < q_{\max}$, the structure factor is suppressed by approximately five orders of magnitude relative to its value outside the constrained region, demonstrating the effectiveness of the generation algorithm.

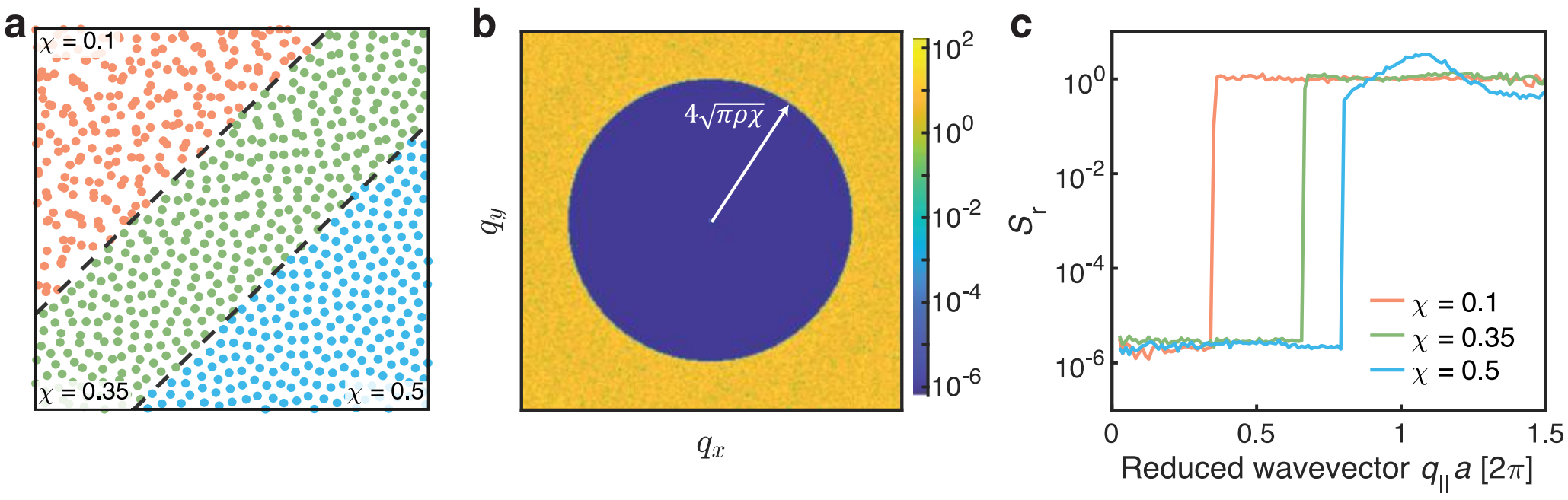

**Figure 2**. **Generation of stealthy hyperuniform patterns. a**, Examples of point patterns with degrees of stealthiness $\chi = 0.1$, 0.35 and 0.5 at fixed point density. Increasing $\chi$ enhances positional order while preserving the absence of long-range periodicity. **b,** Structure factor of a periodic SHU pattern containing $N = 20{,}000$ points with $\chi = 0.35$. (c) Radially averaged structure factors for periodic SHU patterns with $N = 20{,}000$ points and different values of $\chi$. The horizontal axis represents the reduced in-plane wavevector modulus, $q_{||}a$, where $a = 1/\sqrt{\rho}$ is the characteristic inter-particle distance.

Three examples of SHU point patterns with $\chi =$ 0.1, 0.35, and 0.5, each containing $N = 20{,}000$ points, are shown in Figure 2a. Although the optimization algorithm operates in reciprocal space by constraining the structure factor at low wavevectors (Figure 2b), it generates point arrangements that exhibit increasingly pronounced short-range order as $\chi$ increases. This evolution is evident in the emergence of local spatial correlations despite the absence of long-range periodicity.

The corresponding radially averaged structure factors are displayed in Figure 2c. As expected, the extent of the quenching zone increases with $\chi$ in accordance with Eq. (3). Within this region, the structure factor is suppressed by approximately five orders of magnitude, reaching values on the order of $10^{-5}$.

**Metasurfaces fabrication.** The metasurfaces are fabricated using electron-beam lithography in a negative-tone resist, followed by pattern transfer into a 145-nm-thick silicon layer deposited on a glass substrate. More details can be found in [34].

SHU point patterns were generated for degrees of stealthiness $\chi$ ranging from 0.1 to 0.5 and number densities $\rho$ ranging from 2 to 4 μm$^{-2}$. The area of each generated pattern was fixed at $100 \times 100$ μm$^2$. To achieve the target density, the total number of points $N$ was adjusted accordingly. Each pattern was used as a unit cell and tiled into a $3 \times 3$ array, resulting in a metasurface composed of nine replicated copies. The point coordinates were exported in Graphic Data System (GDS) format for lithographic fabrication. All fabricated metasurfaces therefore have square lateral dimensions of 300 μm.

This procedure yielded a comprehensive library of SHU metasurfaces with systematically varied meta-atom size $l$, density $\rho$, and degree of stealthiness $\chi$. Figure 3a shows scanning electron microscope (SEM) images of two representative samples with $\chi = 0.1$ and $\chi = 0.5$, both with $\rho = 2$ μm$^{-2}$ and $l = 100$ nm. The insets in the lower-left corners display the corresponding bright-field optical micrographs. SEM images of the complete set of fabricated metasurfaces are provided in Figure S5.

**Experimental verification of the quenching equation.** Under the independent scattering approximation, the scattered intensity from a metasurface is proportional to the product of the structure factor and the meta-atom form factor [28]. The latter corresponds to the differential scattering cross section of an individual meta-atom. This factorization is analogous to that used in X-ray and neutron scattering, where the scattering pattern results from the interplay between structure and form factors.

The spectrally and angularly resolved scattered light from the metasurfaces is measured using a gonio-spectrometer (Figure S1) illuminated by a slightly focused supercontinuum laser, ensuring nearly uniform illumination across the sample. Owing to the isotropic nature of the metasurfaces, scattering is recorded as a function of the polar detection angle only. After normalization to the diffuse response of a Lambertian reference diffuser, the measurements yield the bidirectional reflectance distribution function (BRDF) [35].

Figure 3b shows the BRDF as a function of the detection polar angle for three SHU metasurfaces with different point densities and degrees of stealthiness under normal-incidence illumination at a wavelength of 480 nm. Diffuse scattering is strongly suppressed within the angular range extending from 0° to $\theta_{\max}$. The extent of this quenching zone depends on both the point density and the degree of stealthiness and is governed by the quenching equation, which, under normal incidence, can be written as

$$\sin(\theta_{\max}) = 2\sqrt{\rho\chi/\pi}\,\lambda, \qquad (4)$$

where $\lambda$ denotes the wavelength. Equation (4) predicts a linear dependence of $\sin(\theta_{\max})$ on $\lambda$, with a slope proportional to $\sqrt{\rho\chi}$.

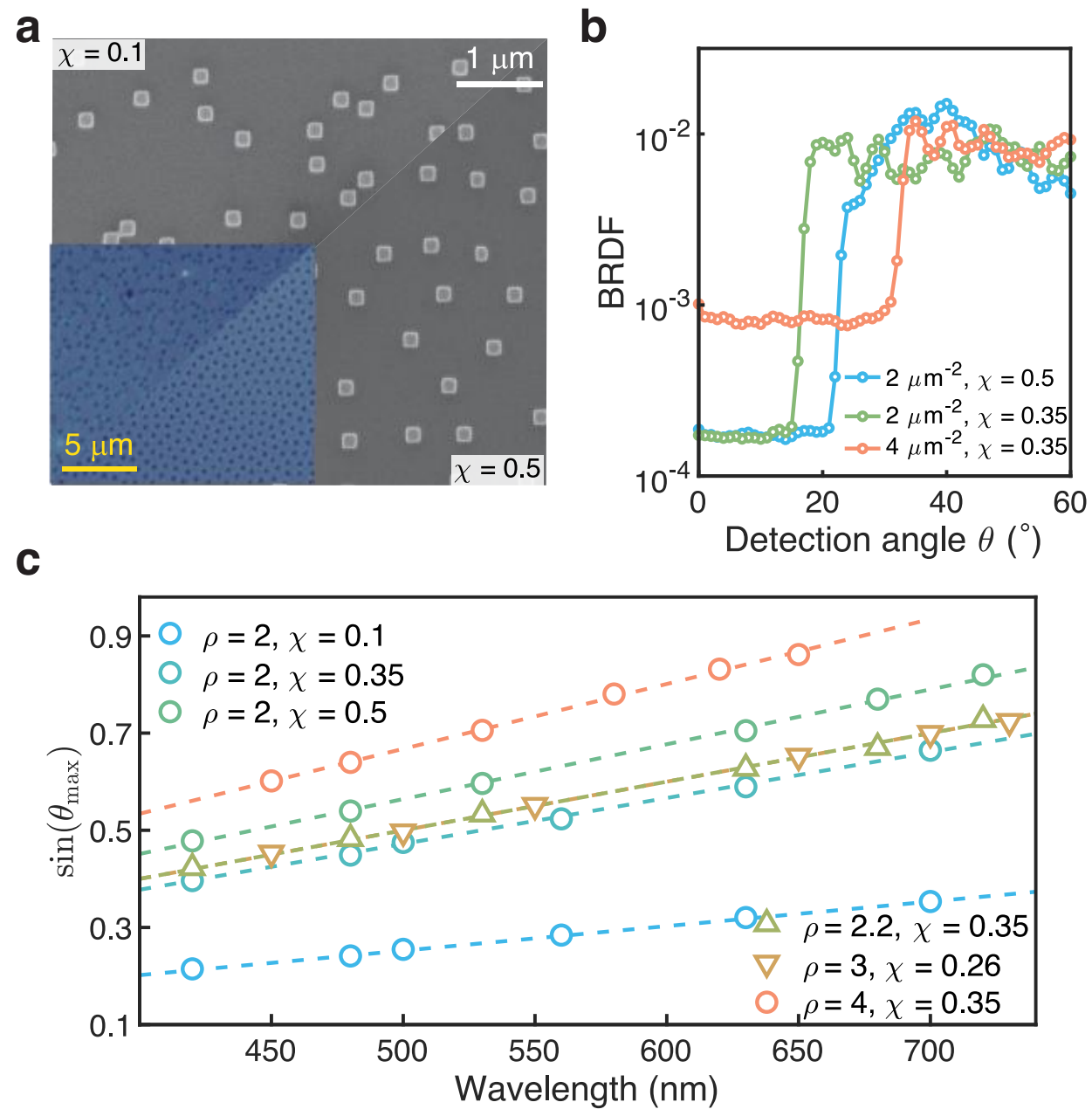

**Figure 3. Verification of the quenching equation. a,** SEM images and corresponding optical microscope images of two samples with $\chi = 0.1$ and $\chi = 0.5$ at a fixed density $\rho = 2\ \mu\mathrm{m}^{-2}$. Scale bars: 1 μm (SEM image) and 5 μm (optical micrographs). **b,** Measured BRDFs as a function of scattering angle $\theta_s$ at $\lambda = 480$ nm for representative metasurfaces. **c,** Experimental verification of the quenching equation for metasurfaces with different values of $\rho$ and $\chi$. Dashed curves show the theoretical prediction of Equation (4), while symbols denote the threshold angles $\theta_{\mathrm{max}}$ extracted from the BRDF measurements.

To validate Equation (4), the transitioning angle $\theta_{\mathrm{max}}$ extracted from measurements performed at various wavelengths for various SHU metasurfaces are shown in Figure 3c. The experimentally extracted values are indicated by circles and triangles, while the dashed curves correspond to Equation (4) using the respective values of $\rho$ and $\chi$. Excellent agreement is observed for all metasurfaces. Notably, for the two metasurfaces indicated by triangles, the products $\rho\chi$ are identical $\rho\chi = \pi/4\ [\mu\mathrm{m}^{-2}]$, resulting in the same slope, and, consequently, overlapping curves. This demonstrates that the quenching zone can be controlled using two independent parameters, $\rho$ and $\chi$, provided their product remains constant.

Together, Equation (3) or Equation (4) provide a practical and intuitive framework for engineering the angular distribution of diffuse scattering, which is of central importance for a wide range of photonic applications, including solar cells [36,37], solid-state lighting devices [7] and colored surfaces in architecture and design [34].

**Quenching efficiency.** Despite the excellent agreement in the angular position of the quenching transition, Figure 3b reveals substantial variations in quenching efficiency among the samples. More importantly, the experimentally measured quenching efficiencies, typically ranging from 10 to 100, are several orders of magnitude smaller than those predicted from the structure factor alone. Additional BRDF measurements performed on metasurfaces with different meta-atom sizes (Figure S3) confirm that the exceptionally large quenching efficiencies predicted theoretically are not attainable under experimental conditions.

## 3. Finite size effect

Although the SHU patterns are generated under periodic boundary conditions, fabrication and measurement constraints inevitably introduce finite-size effects. These may originate from the finite lateral extent of the metasurface or, more fundamentally, from the finite spatial coherence of the incident illumination, which limits the range over which the constructive and destructive interference encoded by the structure factor can develop. This limitation is particularly relevant for applications such as solar cells [36,37], where the spatial coherence length of sunlight is finite and typically on the order of a few tens of micrometers [38].

In finite-size systems, a continuum of wavevectors exists between the discrete reciprocal-lattice vectors $\mathbf{q} = [n_x(2\pi/L), n_y(2\pi/L)]$, where $n_x$ and $n_y$ are relative integers, become accessible. These intermediate wavevectors are not explicitly constrained by the SHU pattern-generation algorithm, and their associated structure factors may therefore be significantly larger than the optimized values obtained at the constrained reciprocal-lattice points.

To quantify the impact of this unconstrained continuum of wavevectors on the quenching efficiency, we first examine SHU patterns containing a small number $N$ of points (Figure 4a–c), which provide a clear visualization of the underlying mechanism, and then summarize the behavior for large $N$ in Figure 4d.

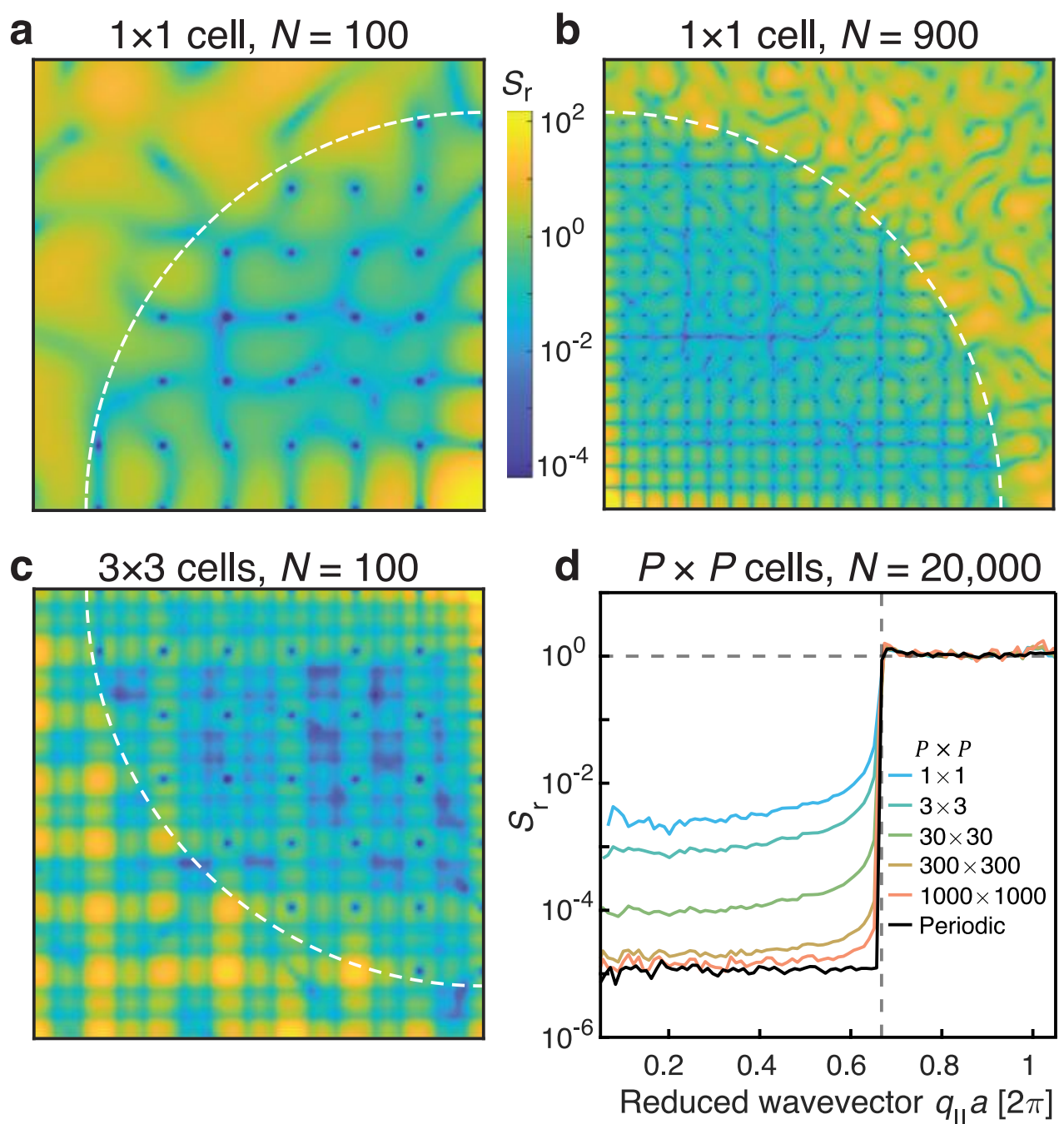

**Figure 4**. **Effect of finite size on the structure factor. a,** Structure factor of a single SHU unit cell containing $N$ =100 points. Strong quenching occurs only at the discrete wavevectors $\mathbf{q} = n_x(2\pi/L)\hat{\mathbf{x}} + n_y(2\pi/L)\hat{\mathbf{y}}$ . **b,** Structure factor of a single unit cell containing $N = 900$ points. Although the constrained wavevectors are more densely distributed, $S_r$ remains significantly larger between them. **c,** Structure factor of a 3×3 tiling of the unit cell shown in panel **a**. **d,** Radially averaged structure factors for $P \times P$ tilings of an SHU unit cell containing $N = 20{,}000$ points, for $P = 1$, 3, 30, 300, 1000 and ∞ (periodic limit). Quenching efficiencies approaching the periodic limit are obtained only for very large values of $P$, typically $P > 300$. All computations are performed for $\chi = 0.35$.

Figure 4a shows the structure factor of a single SHU unit cell for $N = 100$. Even within the quenching zone, delineated by the white dashed arc, the structure factor reaches very small values only at the discrete reciprocal-lattice vectors imposed by periodic boundary conditions. Between these constrained points, the unconstrained continuum of wavevectors exhibits substantially larger values—more than four orders of magnitude higher—which dominate upon momentum-space averaging and therefore strongly reduce the effective quenching efficiency.

Figure 4b shows the corresponding result for a larger unit cell containing $N = 900$ points. In this case, the constrained wavevectors are more densely distributed within the quenching zone. Consequently, the structure factor remains low over a larger fraction of reciprocal space, reducing the contribution of the unconstrained continuum and increasing the average quenching efficiency.

In practice, however, computational limitations prevent arbitrarily large values of $N$ from being used in the generation of a single SHU unit cell. A common alternative is therefore to replicate a smaller optimized unit cell, thereby increasing the overall system size while maintaining a tractable optimization problem.

Consider a finite-size pattern obtained by tiling $P_x \times P_y$ identical replications of a unit cell of size $L \times L$. Denoting by $S_{1\times1}$ the structure factor of a single unit cell, the structure factor of the tiled pattern, $S_{P_x\times P_y}$, is given by [39,32]

$$S_{P_x\times P_y}(\mathbf{q}) = S_{1\times1}(\mathbf{q}) \times \frac{1}{P_x P_y}\left|\frac{\sin(P_x L\mathbf{q}\cdot\hat{\mathbf{x}}/2)}{\sin(L\mathbf{q}\cdot\hat{\mathbf{x}}/2)}\frac{\sin(P_y L\mathbf{q}\cdot\hat{\mathbf{y}}/2)}{\sin(L\mathbf{q}\cdot\hat{\mathbf{y}}/2)}\right|^2. \tag{5}$$

The sinc-like factor corresponds to the interference term arising from the Cartesian tiling of identical unit cells. Equation (5) shows that, as $P_x, P_y$ increase, the structure factor becomes progressively concentrated around the discrete wavevectors $\mathbf{q} = \left(\frac{2\pi n_x}{L}, \frac{2\pi n_y}{L}\right)$, $n_x, n_y \in \mathbb{Z}$, which are precisely the wavevectors constrained by the SHU pattern-generation algorithm through the imposed periodic boundary conditions (Figure S4). In the limit $P_x, P_y \to \infty$, the structure factor vanishes everywhere else in reciprocal space.

Figure 4c shows the structure factor obtained from a 3×3 replication of the SHU unit cell shown in Figure 4a. Compared to Figure 4a, the structure factor at unconstrained wavevectors is substantially larger. As the number of replications increases toward infinity, the sinc-like term converges to a Dirac comb [39], such that only discrete wavevectors contribute to the structure factor, as illustrated in Figure 1b.

The structure factors shown in Figure 4b and Figure 4c correspond to the same total number of points. A direct comparison reveals that the tiled pattern exhibits a more contrasted structure factor and a pronounced anisotropy inherited from the Cartesian replication. Therefore, for a fixed patterned area, generating a single SHU unit cell containing a large number of points is preferable to tiling smaller unit cells, as it yields improved isotropy and a smoother angular response.

Figure 4d summarizes these observations for unit cells containing $N = 20{,}000$ points, corresponding to the patterns used in the experiment. It displays the radially averaged structure factor for metasurfaces composed of $P \times P$ replications, with $P = 1, 3, 30, 300, 1000$ and in the periodic limit $P \to \infty$ (black curve). As expected, the quenching efficiency increases monotonically with $P$. However, efficiencies approaching the periodic limit are reached only for very large values of $P$. For reference, a metasurface composed of 1000×1000 unit cells with $N = 20{,}000$ points at a density of 2 $\mu\text{m}^{-2}$ would occupy an area of 100×100 $\text{mm}^2$.

Fabricating metasurfaces of such dimensions is challenging and achieving illumination with uniform intensity and phase over such an extended area places stringent requirements on the light source. We therefore conclude that the quenching efficiencies predicted for ideal periodic systems are not attainable in practice. Under our experimental conditions, corresponding to $N = 20{,}000$ and $P = 3$, finite-size effects limit the quenching efficiency to approximately $10^3$ (blue-green curve in Figure 4d).

## 4. Effects of nonideal tiling

Disordered stealthy hyperuniform systems are distinguished from naturally occurring, non-stealthy hyperuniform systems by their anomalous suppression of density fluctuations not only in the long-wavelength limit $\mathbf{q} \to 0$ but also over a finite range of wavevectors [12]. This remarkable property originates from globally optimized point configurations, which are typically generated under periodic boundary conditions. In practical implementations, however, large-area metasurfaces are fabricated by replicating finite unit cells, and deviations from the ideal translational continuation imposed by the periodic boundary conditions may occur [17]. Such imperfections modify the structure factor and can degrade the quenching efficiency.

A particularly important source of translational mismatch arises from field-stitching errors during electron-beam lithography. These errors introduce small relative displacements between adjacent writing fields, thereby perturbing the periodic tiling of the SHU unit cell. In this section, we quantify the impact of such stitching errors on the structure factor and the resulting quenching efficiency.

Consider a pattern constructed by replicating a hyperuniform unit cell containing $N_0$ points using $m$ translation vectors $\{\mathbf{R}_1, \ldots, \mathbf{R}_m\}$. The structure factor $S_{\mathrm{tot}}(\mathbf{q}) = \frac{1}{N}\left|\sum_{p=1}^{N} e^{i\mathbf{q}\cdot\mathbf{r}_p}\right|^2$ of the assembled pattern with $N = mN_0$ points is:

$$S_{\mathrm{tot}}(\mathbf{q}) = \frac{1}{N}\left|\sum_{j=1}^{N_0} e^{i\mathbf{q}\cdot(\mathbf{r}_j+\mathbf{R}_1)} + \sum_{j=1}^{N_0} e^{i\mathbf{q}\cdot(\mathbf{r}_j+\mathbf{R}_2)} + \cdots + \sum_{j=1}^{N_0} e^{i\mathbf{q}\cdot(\mathbf{r}_j+\mathbf{R}_m)}\right|^2. \tag{6}$$

Rearranging Eq. (6) gives

$$S_{\mathrm{tot}}(\mathbf{q}) = \frac{1}{N}\left|\left(\sum_{j=1}^{N_0} e^{i\mathbf{q}\cdot\mathbf{r}_j}\right)\left(\sum_{k=1}^{m} e^{i\mathbf{q}\cdot\mathbf{R}_k}\right)\right|^2 = S_{1\times1}(\mathbf{q})S_R(\mathbf{q}), \tag{7}$$

where $S_R(\mathbf{q})$ denotes the structure factor associated with the set of translation vectors $\{\mathbf{R}_1, \ldots, \mathbf{R}_m\}$. For a perfect square tiling, $\mathbf{R}_{mn}^{\mathbf{0}} = mL\,\hat{\mathbf{x}} + nL\,\hat{\mathbf{y}}$, with $m, n \in \mathbb{Z}$, and Eq. (7) reduces to Eq. (5).

In our implementation, the stealthy hyperuniform metasurfaces are fabricated by electron-beam lithography (EBL), with each exposure field corresponding to a single SHU unit cell. Stitching errors between adjacent exposure fields therefore perturb the translation vectors and modify $S_R(\mathbf{q})$. Since the stitching accuracy of our EBL system is specified as 20 nm ± 10 nm (3σ), we consider translational errors ranging from 10 to 50 nm to quantify the impact of nonideal tiling on the structure factor and, ultimately, on the quenching efficiency.

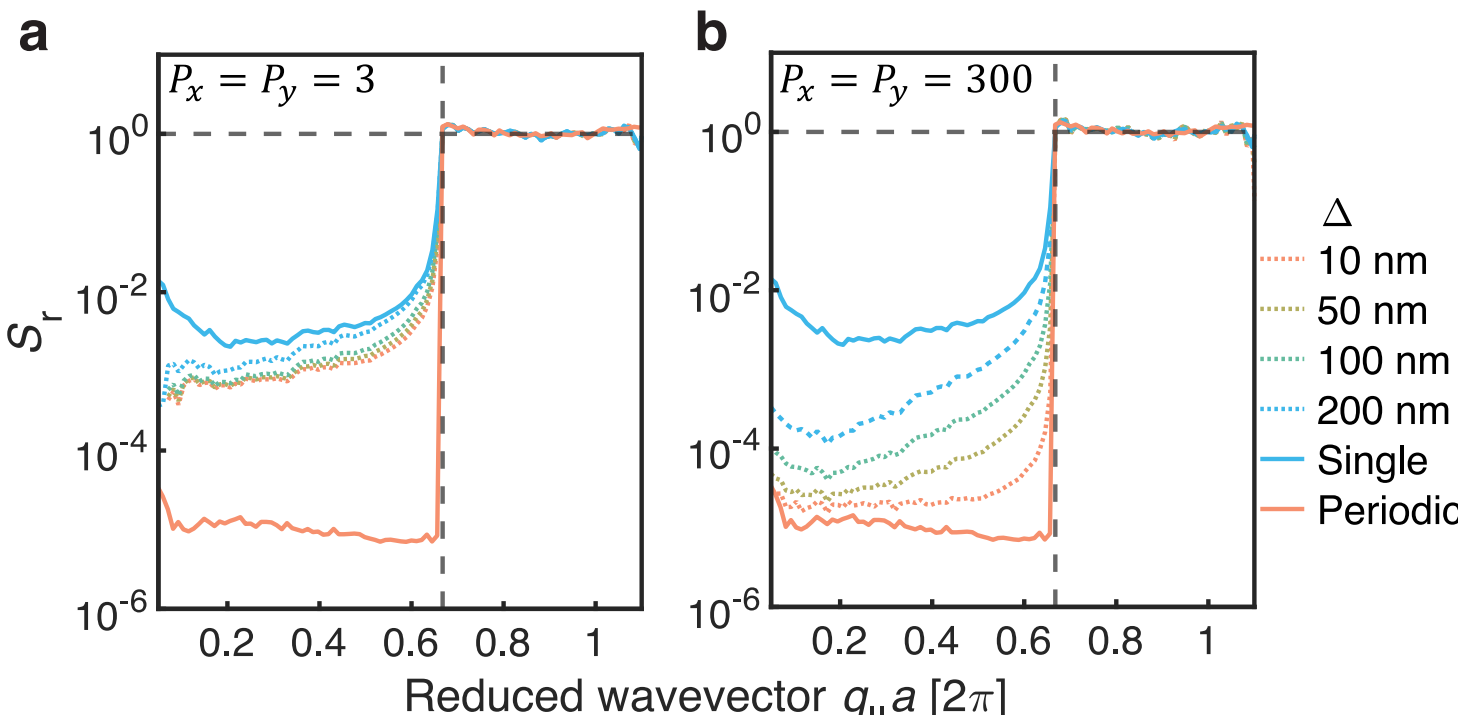

**Figure 5. Effect of stitching errors on the structure factor.** Radial averaged factors of stitched patterns assembled from a 20,000-point SHU unit cell. **a,** 3×3 tiled pattern under varying stitching errors. **b,** 300×300 stitched pattern for different levels of stitching error. In all calculations, stitching errors are modeled by adding independent random displacements to the ideal translation vectors $\mathbf{R}^{(0)}$, with Cartesian components drawn from Gaussian distributions of standard deviation $\sigma(\Delta R_x) = \sigma(\Delta R_y) = \Delta$ in both directions. Each curve is averaged over five independent random realizations of the translation-vector set. The SHU unit cell has a point density $\rho = 2\ \mu\mathrm{m}^{-2}$ and a degree of stealthiness $\chi = 0.35$.

To incorporate stitching errors, the ideal translation vectors are perturbed according to $\mathbf{R}_k = \mathbf{R}_k^{\mathbf{0}} + \mathbf{\Delta R}_k$ , where $\mathbf{\Delta R}_k$ represents the write-field alignment error. The Cartesian components of $\mathbf{\Delta R}_k$ are independently sampled from Gaussian distributions with standard deviation Δ along both directions. To assess their impact, different levels of stitching error are introduced into both 3×3 and 300×300 tiled patterns.

Figure 5a shows the resulting structure factor for the 3×3 tiled pattern used in the experiment. Owing to the finite-size effects discussed above, the structure factor in the absence of stitching errors is already close to that of a single unit cell, and stitching errors within the typical EBL range induce only minor modifications. In contrast, finite-size effects are much weaker in the 300×300 tiled pattern, which therefore provides a closer approximation to the infinite-periodic limit. As shown in Figure 5b, increasing stitching errors progressively drive the structure factor away from the periodic limit and toward the single-unit-cell limit.

These results show that stitching errors hinder the convergence of large tiled structures toward the infinite-periodic limit. Nevertheless, even when realistic stitching errors are included, the resulting changes in the structure factor remain too small to explain the large discrepancy between theoretical and experimental quenching efficiencies. The quenching efficiencies predicted from the structure factor alone remain substantially larger than the maximum value measured experimentally, approximately 70 (Figure S3). Additional limiting mechanisms must therefore be considered.

## 5. Effect of polydispersity

Electron-beam lithography enables meta-atoms to be positioned with high accuracy, typically at the nanometer scale. Nevertheless, fabrication-induced variations in the etching process introduce small but finite deviations in shape and size, leading to meta-atom-to-meta-atom variations in the scattered electromagnetic field. In the presence of such variations, Equation (2), which assumes identical scattering properties for all meta-atoms, is no longer valid and the more general formulation of Equation (1) must be considered. Under the independent-scattering approximation, the effect of polydispersity can be incorporated at the level of the individual far-field response of every meta-atom.

Consider a collection of $N$ meta-atoms, each with a different size or shape symbolically denoted by $l^{(p)}$, $p = 1, 2 \dots N$. The collection is illuminated by an incident plane wave. Using a near-to-far-field transformation, we compute for each meta-atom the amplitude of the plane-wave component scattered in the direction $\mathbf{k}_\mathrm{s}$ with polarization $\hat{\mathbf{e}}_\mathrm{s}$ [40]. This complex amplitude is denoted by $\alpha^{(p)}(\mathbf{k}_\mathrm{s}, \mathbf{k}_\mathrm{i}, \hat{\mathbf{e}}_\mathrm{s})$. Similarly, we define $\alpha^{(0)}(\mathbf{k}_\mathrm{s}, \mathbf{k}_\mathrm{i}, \hat{\mathbf{e}}_\mathrm{s})$ as the corresponding amplitude for a reference meta-atom with nominal size $l_0$.

To quantify the effect of fabrication-induced variations, we introduce the normalized scattering amplitude $\delta^{(p)}(\mathbf{k}_\mathrm{s}, \mathbf{k}_\mathrm{i}, \hat{\mathbf{e}}_\mathrm{s}) = \alpha^{(p)}(\mathbf{k}_\mathrm{s}, \mathbf{k}_\mathrm{i}, \hat{\mathbf{e}}_\mathrm{s}) / \alpha^{(0)}(\mathbf{k}_\mathrm{s}, \mathbf{k}_\mathrm{i}, \hat{\mathbf{e}}_\mathrm{s})$. The quantity $\delta^{(p)}$ is a complex weight that measures the deviation of the scattering response of meta-atom $p$ from that of the reference structure. In the absence of shape and size variations, $\delta^{(p)} \to 1$.

Including these fluctuations, the scattered amplitude of the collection in the direction $\alpha_c$ in the $\mathbf{k}_\mathrm{s}$ direction and polarization $\hat{\mathbf{e}}_\mathrm{s}$ becomes

$$\alpha_c(\mathbf{k}_\mathrm{s}, \mathbf{k}_\mathrm{i}, \hat{\mathbf{e}}_\mathrm{s}) = \alpha^{(0)}(\mathbf{k}_\mathrm{s}, \mathbf{k}_\mathrm{i}, \hat{\mathbf{e}}_\mathrm{s}) \sum_{p=1}^{N} \delta^{(p)}(\mathbf{k}_\mathrm{s}, \mathbf{k}_\mathrm{i}, \hat{\mathbf{e}}_\mathrm{s}) \exp\left[-i\mathbf{q}_\parallel \cdot \mathbf{r}_{p,\parallel}\right]. \tag{8}$$

Relative to the monodisperse case, polydispersity is therefore incorporated through a set of complex weights $\delta^{(p)}$ fluctuating around unity. This naturally leads to the definition of a modified structure factor $S_\mathrm{poly}$ [41,42],

$$S_\mathrm{poly}(\mathbf{k}_\mathrm{s}, \mathbf{k}_\mathrm{i}, \hat{\mathbf{e}}_\mathrm{s}) = \frac{1}{N}\left|\sum_{p=1}^{N} \delta^{(p)}(\mathbf{k}_\mathrm{s}, \mathbf{k}_\mathrm{i}, \hat{\mathbf{e}}_\mathrm{s}) \exp\left[-i\mathbf{q}_\parallel \cdot \mathbf{r}_{p,\parallel}\right]\right|^2, \tag{9}$$

which accounts for both positional disorder and scatterer-to-scatterer variations. For a given realization of the positions $\mathbf{r}_{p,\parallel}$, this quantity can be statistically averaged over the distribution of meta-atom geometries.

Although we refer to $S_{\text{poly}}$ as a structure factor, it is not determined solely by positional correlations. In contrast to the conventional structure factor, it also incorporates fluctuations in the scattering amplitudes of the individual meta-atoms. Factoring out the mean form factor in this manner provides a quantity that can be directly used for analyzing the experimental quenching efficiency.

To investigate the effect of polydispersity, the statistical distribution of the lateral size of the Si nanoboxes is extracted from SEM images. Analysis of approximately 212 nanoboxes yields an average lateral size of $l_0 = 111.7 \pm 1.8$ nm (Figure 6a). Based on this distribution, the near field scattered under normal incidence at $\lambda = 532$ nm is computed using the open-source freeware RETOP [40] for a set of nanoboxes with lateral sizes varying in steps of 1 nm. At this wavelength, the refractive index of silicon is taken as $n_{\text{Si}} = 4.3 + 0.16i$ , based on experimentally measured values. In all computations, the phase origin is chosen at the substrate interface, at the center of the bottom face of each nanobox.

Because we consider in-plane scattering directions only (the azimuthal angle is null) and a TE polarized incident plane wave, owing to symmetry, the scattered plane wave is also TE polarized for all scattering directions. Figure S6 shows the phase and modulus of $\delta^{(p)}$, normalized to the reference size $l_0$, for all scattering directions considered. Because the size variations are small and the dependence on the scattering direction is nearly identical in all cases, only the results obtained for a polar angle of 30° in Figure 6b.

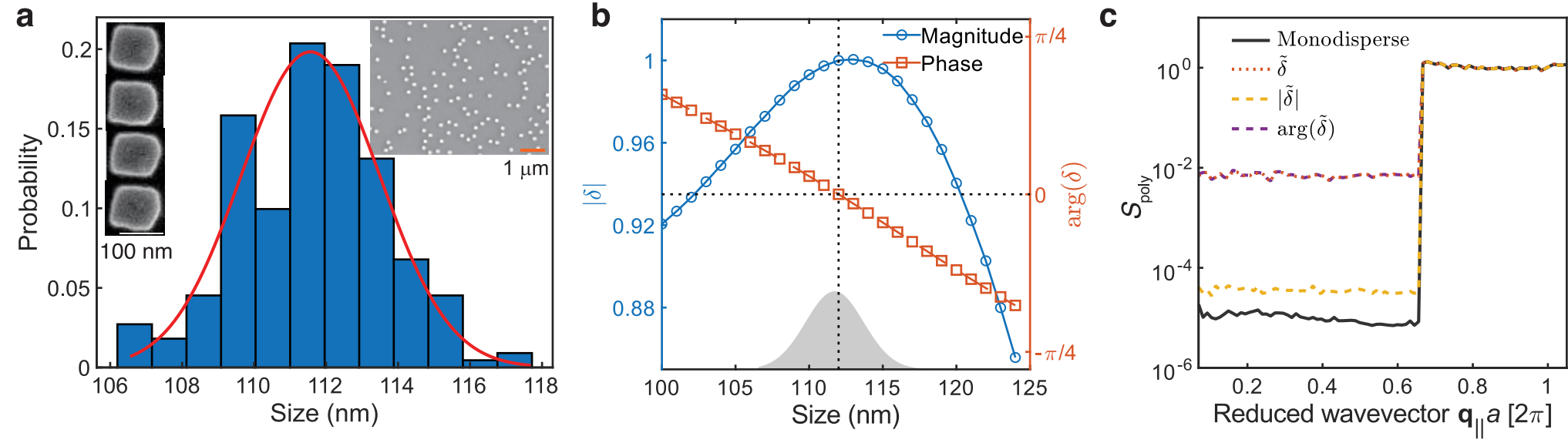

**Figure 6**. **Effect of polydispersity on the structure factor. a**, Statistical distribution of nanobox sizes. Inset: SEM images; scale bars: 100 nm (left), 1 μm (right). **b**, Simulated scattered-field fluctuations $\delta^{(p)}$ for nanoboxes of different sizes, with amplitudes normalized to the complex amplitude of the average-size meta-atom. Left: magnitude; Right: phase. **c**, Polydisperse structure factor $S_{\text{poly}}$ obtained using Equation (9).

Based on the experimentally measured size distribution, lateral sizes are randomly assigned to the Si nanoboxes to compute the modified structure factor $S_{\text{poly}}$ for $N = 20{,}000$ points and $\chi = 0.35$. Figure 6c compares the structure factors obtained for the monodisperse and polydisperse cases. Polydispersity dramatically reduces the quenching efficiency, from $10^5$ to $10^2$.

To identify the origin of this reduction, we separately evaluate the contributions of amplitude and phase fluctuations. As shown in Figure 6c, phase fluctuations are responsible for most of the observed decrease. Indeed, the phase of $\delta^{(p)}$ varies rapidly in the vicinity of the reference size $l_0$. We attribute this behavior to a resonance of the meta-atom. This interpretation is further supported by the non-monotonic dependence of $|\delta^l|$ on size, shown in Figure 6b, which exhibits a characteristic increase followed by a decrease across the resonance.

In practice, fabricated nanoboxes exhibit geometrical variations beyond lateral size dispersion, including corner rounding and edge roughness. To keep the analysis tractable, only lateral size variations are considered here, as they constitute the dominant source of fluctuations in the scattering response of subwavelength Mie-resonant dielectric nanoparticles. Additional simulations show that the fluctuations induced by corner rounding are significantly smaller than those resulting from size variations (see Figure S8).

We conclude that finite-size effects and polydispersity both contribute to reducing the quenching efficiency from $10^5$ to $10^2$. However, these mechanisms alone cannot explain two key experimental observations. First, the measured quenching efficiency, typically of the order of 10, remains substantially lower than the values predicted when finite-size and polydispersity effects are included. Second, the experimentally observed quenching efficiency decreases as the meta-atom density increases, whereas the independent-scattering approximation predicts the opposite trend because a higher density corresponds to a larger number of scatterers. These discrepancies indicate that additional physical mechanisms play a dominant role in limiting the performance of SHU metasurfaces.

### 6. Effect of multiple-scattering

To investigate the impact of multiple scattering on the quenching efficiency, we perform full-wave electromagnetic simulations using the open-source software RETICOLO [43], which implements the rigorous coupled-wave analysis (RCWA) method [44]. All simulations are performed at $\lambda$ = 532 nm under normal incidence.

Owing to computational constraints, SHU patterns containing only $N = 26$ points are generated. For the simulations, Si nanoboxes with lateral dimensions of 100 nm and a height of 145 nm, matching the experimental structures, are positioned at the SHU coordinates on a silica substrate (Figures S8 and S9).

Figure 7a and 7b show the statistically averaged computed diffraction efficiencies of the reflected diffraction orders of SHU metasurfaces with density $\rho = 3$ µm$^{-2}$ and $\chi = 0.35$. Each diffraction order is represented by a square, whose position is determined by the normalized in-plane wavevector components $k_x/k_0$ and $k_y/k_0$. The supercell size is $L = \sqrt{N/\rho} \approx 5.53\,\lambda$. The white dashed circles indicate the quenching boundary predicted by Equation (4), while the black dashed circles represent the air light line, $k_x^2 + k_y^2 = k_0^2$.

The diffraction efficiencies computed for the SHU supercells are shown in Figure 7b. The results are averaged over 40 statistically independent realizations generated with the SHU algorithm. For comparison, we also evaluate the diffraction efficiencies within the framework of the ISA, as shown in Figure 7a.

To isolate the effect of multiple scattering, the ISA calculations are performed using exactly the same numerical parameters as the full-wave RCWA simulations: identical incident plane wave, supercell size, and number of Fourier harmonics retained in the expansion. We first compute the electromagnetic field scattered by a single meta-atom located at the center of the supercell into all diffraction orders. The scattered field of the SHU supercell is then obtained analytically by summing the contributions from all meta-atoms, each weighted by the phase factor $e^{i(\mathbf{k}_\mathrm{i}-\mathbf{k}_\mathrm{S})\cdot r_{p,\parallel}}$. *In this framework, the meta-atom contributions differ only through their in-plane positions* $r_{p,\parallel}$. Diffraction efficiencies are subsequently calculated from the corresponding Poynting vectors for every diffraction order. The procedure is repeated for each of the 40 independent SHU realizations.

In both Figures 7a and 7b, the abrupt transition predicted by the quenching equation is clearly visible. However, the quenching efficiency obtained from the full-wave RCWA simulations is reduced by approximately five orders of magnitude compared with the ISA prediction.

This discrepancy cannot be attributed to polarization effects, angular dependence, or the vector nature of electromagnetic scattering, since these aspects are treated identically in both calculations. The only difference is that the ISA neglects recurrent electromagnetic interactions between meta-atoms, whereas these interactions are fully included in the RCWA simulations. The large discrepancy between Figures 7a and 7b therefore demonstrates that multiple scattering plays a dominant role in determining the quenching efficiency of SHU metasurfaces.

One may question whether a numerical supercell containing only $N = 26$ meta-atoms can fully reproduce the isotropy and dense reciprocal-space sampling of the experimental structures, which contain up to $N = 20\,000$ points. Consequently, it is legitimate to ask whether multiple scattering has a comparable impact in the fabricated samples.

A direct numerical verification is currently beyond our computational capabilities. In addition, the smallest non-zero wavevector accessible in the supercell approach is $2\pi/L$, such that the low-**q** regime cannot be directly investigated. However, the quenching efficiency considered here is evaluated both experimentally and theoretically as the ratio of the averaged diffraction efficiencies inside and outside the quenching wavevector range, making it a meaningful statistical measure despite the limited low-**q** sampling. Nevertheless, the RCWA simulations fully account for recurrent electromagnetic interactions among the meta-atoms. These interactions occur both at short range within the supercell and at long range through the periodic repetition of the supercell imposed by the simulation. Furthermore, averaging over 40 statistically independent SHU realizations reduces the influence of a particular configuration and provides a representative estimate of the multiple-scattering effects expected in larger hyperuniform systems with the same density $\rho$ and stealthiness parameter $\chi$.

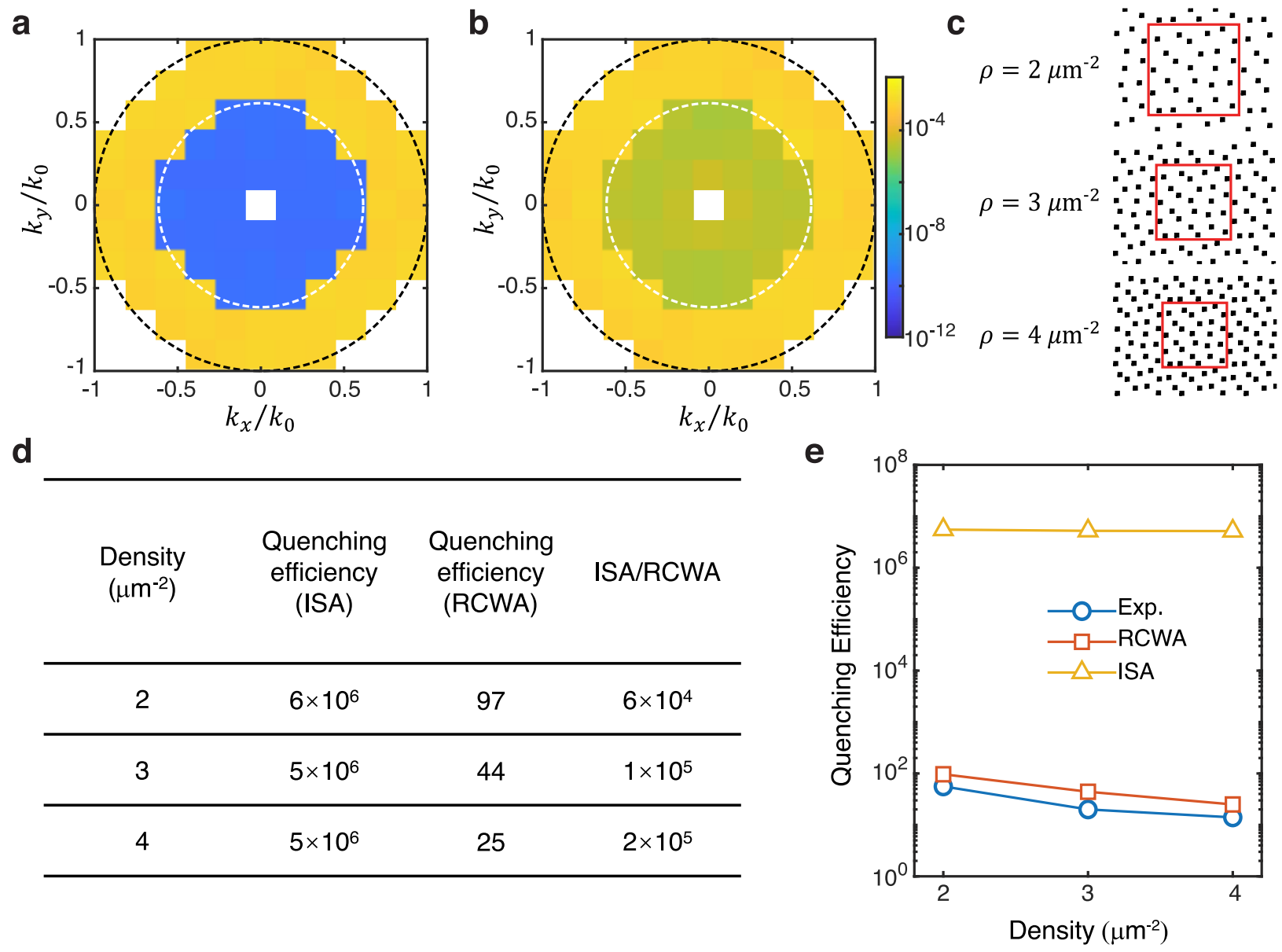

| Density ($\mu m^{-2}$) | Quenching efficiency (ISA) | Quenching efficiency (RCWA) | ISA/RCWA |
|---|---|---|---|
| 2 | $6\times10^6$ | 97 | $6\times10^4$ |
| 3 | $5\times10^6$ | 44 | $1\times10^5$ |
| 4 | $5\times10^6$ | 25 | $2\times10^5$ |

**Figure 7**. **Impact of multiple scattering on quenching efficiency. a**, Reflected diffraction efficiencies in all orders of a 2D grating with a SHU unit cell containing $N = 26$ Si identical nanoboxes computed under the independent scattering approximation (ISA) for $\rho = 3\ \mu m^{-2}$ and $\chi = 0.35$. Each diffraction order is represented by a square box. White boxes correspond to evanescent orders with in-plane wavenumber larger than the free space wavenumber $k_0 = 2\pi/\lambda$. The diffraction efficiency in the (0,0) order (specular light) is also represented as a white box. **b**, RCWA results (multiple scattering is taken into account) for the same SHU pattern as in (a). **c**, Schematic illustration of unit cells for SHU patterns with identical arrangements $\chi = 0.35$ at different densities. **d**, Calculated quenching efficiency of SHU metasurfaces as a function of density. As the density increases, the deviation from the ISA prediction becomes more pronounced. **e**, Comparison between quenching efficiencies obtained from experiment, RCWA simulations, and ISA predictions. Data in panels **d** and **e** are obtained by averaging over 40 statistically independent realizations of patterns generated with the SHU generator. See Figures S8-S11 for more details, including convergence tests on the accuracy of the computed data.

To further support our analysis, we perform additional checks to verify that the trends observed experimentally for large systems are also reproduced numerically in supercells of different sizes accessible to full-wave simulations (see Figure S14).

From a physical standpoint, multiple-scattering effects are expected to become stronger as the meta-atom density increases. This trend is indeed observed experimentally. Figure S3 shows the measured bidirectional reflectance distribution functions (BRDFs) of SHU metasurfaces with identical stealthiness parameter $\chi = 0.35$ and densities $\rho = 2$, 3, and 4 $\mu m^{-2}$.

To investigate this effect numerically, we uniformly rescale the coordinates of a fixed SHU pattern containing $N = 26$ points to generate metasurfaces with different densities, as illustrated in Figure 7c. This procedure preserves the structure factor as a function of the reduced in-plane wavevector $q_{\parallel}/\sqrt{\rho}$. RCWA simulations are then performed for each density using the same underlying SHU configuration.

Figure 7d compares the quenching efficiencies obtained from the RCWA simulations with those predicted by the ISA for the three densities considered. As expected, the RCWA calculations yield quenching efficiencies that are approximately five orders of magnitude lower than the ISA predictions, with the discrepancy becoming more pronounced as the density increases. In contrast, the ISA predicts a quenching efficiency that is nearly independent of density.

Figure 7e summarizes the experimental measurements together with the corresponding RCWA simulations and independent-scattering-approximation predictions for SHU metasurfaces at various densities. Although the exact magnitude of the effect may depend on the system size, these results consistently demonstrate that multiple scattering imposes an intrinsic limitation on the achievable quenching efficiency of SHU metasurfaces.

Importantly, this limitation is expected to persist in the thermodynamic limit [45], where the system size tends to infinity and all meta-atoms are identical. In this limit, the structure-factor-based predictions of SHU pattern generators break down because they neglect the unavoidable electromagnetic coupling between neighboring scatterers. Consequently, the quenching efficiency predicted from the structure factor alone can substantially overestimate the performance of real electromagnetic metasurfaces.

## 7. Outlook

In SHU design, the key challenge is the accurate positioning of points to control interference in momentum space. State-of-the-art SHU algorithms enable the engineering of either destructive or constructive interference with high accuracy over a broad range of momentum patterns by positioning millions of points.

We employed these algorithms to design periodic optical metasurfaces with SHU unit cells. The metasurfaces were fabricated using electron-beam lithography and reactive ion etching, resulting in SHU arrays of silicon nanoboxes on a silica substrate.

Our measurements reveal a pronounced suppression of diffuse scattering around the specular direction, with an angle-dependent threshold that systematically follows the quenching condition predicted by Equation (3). However, the experimentally observed suppression remains significantly weaker than that predicted from structure-factor calculations.

We systematically investigated the possible origins of this discrepancy. First, achieving very high suppression requires extremely large SHU patterns, on the order of $10^{10}$ points, corresponding to areas approaching $10^4$ $mm^2$ for a density of 2 $\mu m^{-2}$. Second, polydispersity constitutes a major limitation. Even for simple meta-atoms such as silicon nanoboxes, fabrication-induced size variations generate additional diffuse scattering within the quenching region. Within realistic fabrication tolerances, phase fluctuations are found to be considerably more detrimental than amplitude fluctuations. Third, multiple scattering strongly limits the achievable suppression, even at densities commonly regarded as low in photonic applications, such as 2 $\mu m^{-2}$. Finally, stitching errors introduced during electron-beam lithography are shown to have a negligible impact on the measured performance.

These results suggest limited opportunities for substantially improving the performance of SHU devices within the current design paradigm. Multiple-scattering effects may be mitigated through the use of specifically engineered meta-atoms [46], while highly ordered structures with large $\chi$ values provide a robust alternative, as demonstrated here. The impact of polydispersity can be reduced by employing non-resonant meta-atoms, although this inevitably decreases the diffuse-scattering signal. From an algorithmic perspective, it may be advantageous to develop optimization strategies that target moderate suppression levels within constrained regions of momentum space, potentially yielding more robust performance elsewhere.

In our best experimental realization, a suppression factor of 70 was achieved, with typical values ranging between 10 and 30. Such performance is already sufficient for several light-harvesting applications, including solar-energy technologies [36,37]. More demanding applications, such as transparent displays, will require substantially higher suppression levels because of the high sensitivity of human vision to residual optical noise. Nevertheless, the present work establishes a quantitative framework for identifying and ranking the mechanisms that limit the performance of SHU devices, providing clear guidelines for future designs.

## Methods

**Sample fabrication.** The metasurfaces were fabricated by etching structures into a 145 nm-thick polycrystalline silicon (Si-poly) layer. First, Si-poly was deposited on both sides of a 4-inch fused silica (FS) wafer by LPCVD at 605 °C. The Si-poly on one side of the wafer was then removed using fluorine-based reactive ion etching (F-RIE). The sample was subsequently coated with a 160 nm-thick layer of negative resist (maN2405, Micro Resist Technology) and a 40 nm-thick conductive layer (ELECTRA92, AllResist) to suppress charging during electron-beam lithography (EBL).

Patterns were exposed using EBL with a beam energy of 20 keV, a current of 135 pA, a step size of 4 nm, and a nominal dose of 150 µC/cm². The resist was developed in MF-CD-26 (Microposit) at 20 °C for 50 s, rinsed with deionized water for 1 minute, and dried with nitrogen. The negative maN resist patterns were then transferred into the Si-poly layer using F-RIE, and the remaining resist mask was removed by oxygen plasma. No proximity effect correction (PEC) was applied during electron-beam lithography. Since the proximity effect at 20 keV extends over a region of approximately 4 µm in diameter and the local density of the hyperuniform patterns is statistically uniform within this window range, its influence is expected to be nearly homogeneous across the sample. After resist development and reactive-ion etching, the actual nanobox dimensions were determined from SEM images and were found to deviate slightly from the design values.

**Optical Characterization.** Images of the samples were acquired in bright-field mode using a Nikon 100× objective and a camera (Imaging Source, DFK 33UX264). BRDF measurements are conducted using a custom-built gonio-spectrometer setup equipped with a supercontinuum laser and a 1 mm-diameter optical fiber connected to a spectrometer. The incidence ($\theta_i$) and scattering ($\theta_s, \phi_s$) angles are precisely controlled by three stepper motor rotation stages (Newport URS75, URS150, and SR50CC). As shown in Figure S1, the scattered light was collected slightly above the plane of incidence by a 1 mm diameter optical fiber connected to a spectrometer (Ocean Insight, HDX). The BRDF values were calibrated by measuring a Lambertian reflector with a reflectance of 99.99%, Spectralon (Labsphere SRS-99-010), using the same experimental setup.

**Electromagnetic computational results.** To compute the radiation diagrams of a single Si meta-atom shown in Figure 6, we first used COMSOL Multiphysics 6.3 to calculate the near-field electromagnetic response of an isolated silicon nanobox placed on a silica substrate. The refractive index of silicon was taken from experimentally measured values, $n_{\mathrm{Si}} = 4.3 + 0.16i$, while the refractive index of the silica substrate was set to 1.46. The open-source software RETOP [40] was then employed to perform the near-to-far-field transformation, yielding the plane-wave expansion of the far-field radiation for Si nanoboxes of different sizes.

For the results presented in Figure 7, each supercell consists of 26 Si meta-atoms arranged on a silica substrate, with air as the superstrate. The diffraction efficiencies in all orders are computed using the open-source software RETICOLO [43]. To ensure numerical convergence, at least $81 \times 81$ Fourier harmonics are retained in the calculations. Simulations are performed under normal incidence, and the total diffraction efficiency is obtained by averaging the TE and TM polarization contributions for each diffraction order.

For the corresponding independent-scattering-approximation calculations, a single Si nanobox is placed at the center of a supercell of identical size. The diffraction efficiencies of all orders are again computed using RETICOLO, with the same number of retained Fourier harmonics to avoid numerical discrepancies. These single-particle diffraction efficiencies are then multiplied by the numerically computed structure factor of each SHU pattern, as defined in Equation (2). This approach allows the ISA predictions to be obtained very efficiently while maintaining the same level of numerical accuracy as the full RCWA calculations.

## Acknowledgements

PL acknowledges financial support from the Grand Research Program LIGHT Idex of Bordeaux University and the European Research Council Advanced grant (Project UNSEEN No. 101097856). The authors thanks Philippe Teulat for his repeated help in developing the goniospectrometer setup. Sample fabrication was supported by LAAS-CNRS micro and nanotechnologies platform, a member of the RENATECH French national network.

## Competing interests

The authors declare no competing interests.